\documentclass[twocolumn,preprintnumbers,amsmath,floatfix,prl,aps,superscriptaddress]{revtex4}

\usepackage{graphicx}
\usepackage{ifthen}
\usepackage{ifpdf}
\usepackage{color}
\definecolor{red}{rgb}{1.,0.,0.}

\usepackage{latexsym}
\usepackage{amssymb}
\usepackage{bm}
\newcommand{\half}{\mbox{\small $\frac{1}{2}$}}

\newcommand{\eexp}{\mbox{e}^}

\newcommand{\beq}[1]{\begin{eqnarray}\ifthenelse{#1=-1}{\nonumber}
{\ifthenelse{#1=0}{}{\label{e#1}}}}
\newcommand{\eeq}{\end{eqnarray}}
\newcommand{\be}[1]{\begin{eqnarray}}
\newcommand{\ee}{\end{eqnarray}}

\newcommand{\hide}[1]{}

\begin{document}

\title{A model for the ESR-STM phenomenon}
\author{Alvaro Caso}
\affiliation{Departamento de F\'{\i}sica, Universidad de Buenos Aires, Ciudad
  Universitaria Pabell\'on I, (1428) Buenos Aires, Argentina}
\affiliation{Max-Planck-Institut f\"ur Physik komplexer Systeme, N\"othnitzer
  Str.~38, D-01187 Dresden, Germany} 

\author{Baruch Horovitz}
\affiliation{Department of Physics, Ben Gurion University,
Beer Sheva 84105 Israel}

\author{Liliana Arrachea}
\affiliation{Departamento de F\'{\i}sica, Universidad de Buenos Aires, Ciudad
  Universitaria Pabell\'on I, (1428) Buenos Aires, Argentina}

\begin{abstract}
  We propose a model to account for the observed ESR-like signal at the Larmor
  frequency in the current noise STM experiments identifying spin centers on
  various substrates.  The theoretical understanding of this phenomenon, which
  allows for single spin detection on surfaces at room temperature, is not
  settled for the experimentally relevant case that the tip and substrate are
  not spin polarized.  Our model is based on a direct tip-substrate tunneling
  in parallel with a current flowing via the spin states. We find a sharp
  signal at the Larmor frequency even at high temperatures, in good agreement
  with experimental data. We also evaluate the noise in presence of an ac
  field near resonance and predict splitting of the signal by the Rabi
  frequency.
\end{abstract}


\maketitle

Observing and manipulating single spins is of considerable interest in quantum
information. A particularly promising method of detecting a single spin on a
surface is by using a Scanning Tunneling Microscope (STM) \cite{review}. The
technique has been initiated and developed by Y. Manassen and various
collaborators and has the appealing feature of being useful at ambient
conditions, in contrast to other techniques operating at very low temperatures
\cite{lowt}. It is based on monitoring the noise, i.e. the current-current
correlations, in the STM current and observing a signal at the expected Larmor
frequency, similar to an Electron Spin Resonance (ESR) experiment, except that
here no oscillating field is applied. The frequency of the signal varies
linearly with the applied magnetic field, confirming that the STM in fact
detects an isolated spin on the surface. This phenomenon was demonstrated on
oxidized Si(111) surface \cite{manassen1,manassen2}. Afterwards it was also
observed in Fe atoms \cite{manassen3} on Si(111) as well as on a variety of
organic molecules on a graphite surface \cite{durkan} and on Au(111) surfaces
\cite{messina,mannini,mugnaini}. Recent extensions have resolved two resonance
peaks on oxidized Si(111) $7\times 7$ surface corresponding to site specific
$g$ factors \cite{komeda,sainoo} and also enabled the observation of the
hyperfine coupling \cite{review,manassen4}.

The theoretical understanding of the ESR-STM effect is not settled
\cite{review}. The emergence of a finite frequency in a steady state
stationary situation is a non-trivial phenomenon. An obvious mechanism for
coupling the charge current to the spin precession is spin-orbit coupling
\cite{mozyrsky}. It was shown, that an ESR signal is present in the noise of
systems with spin-orbit coupling only when the leads are polarized, either for
a strong Coulomb interaction \cite{bulaevskii,gurvitz} or for the
non-interacting case \cite{gurvitz}.  This was shown even in linear response
\cite{entin}. However, it was found that the signal vanishes when the lead
polarization vanishes or with parallel polarizations. In the experiments the
leads are very weakly polarized by the magnetic field with polarization
parallel to that of the localized spin, hence the ESR signal vanishes within
these models \cite{bulaevskii,gurvitz}.  It was argued that an effective spin
polarization is realized as a fluctuation effect either for a small number of
electrons that pass the localized spin in one cycle \cite{balatsky} or due to
1/f magnetic noise of the tunneling current \cite{manassen5}.  It was further
shown that spin-orbit coupling in an asymmetric dot can yield an oscillating
electric dipole, possibly affecting the STM current \cite{levitov}.

In the present work we start by showing that the problem of tunneling via spin
states in the presence of spin-orbit interaction has strictly no resonance
signal; the presence of an electric dipole coupling does not change this
conclusion. We then study a model in which an additional direct coupling
between the dot and the reservoir is included. Our model is motivated by
studies of quantum dots with spin-orbit \cite{lopez} and by STM studies of a
two-impurity Kondo system that shows a significant direct coupling between the
tip and substrate states \cite{bork,hamad}.  We find that the interference of
the direct current and the one via the spin does show an ESR signal in the
noise, which increases with the direct coupling. This feature is consistent
with the unusual non-monotonic contour plot presented in Refs.
\onlinecite{manassen1,manassen2}, i.e. the signal is maximized when the STM
tip is not directly on the spin center but slightly ($\sim 1$nm) away, so as
to maximize an overlap with a surface state of the substrate. The signal
intensity relative to the background is small, yet it is sharp even at
temperatures much higher than the ESR frequency; the consistency of this
behavior with the experimental data is discussed below. Finally, we also
evaluate the noise in presence of an ac field near resonance and predict
splitting of the signal.

The setup we consider consists of a molecule $M$ with spin $1/2$ on a metallic
substrate and also in contact with an STM tip. The tip and
the substrate define two electron reservoirs $T,S$. The two states of
the molecule, associated to the two spin orientations of the molecule have energies
separated by a Zeeman splitting due to an external magnetic field ${\bf B}= B
\hat{e}_z$, and are coupled by tunneling processes to the two reservoirs. We
assume, in addition, spin-orbit (SO) coupling, which renders these tunneling
terms spin dependent, in particular allowing for tunneling with spin-flip.
The Hamiltonian is \beq{01} {\cal H}= \sum_{\alpha=T, S } {\cal H}_{\alpha}+
{\cal H}_{M} + \sum_{\alpha=T, S} {\cal H}_{M,\alpha}.  \ee The first term,
with $ {\cal H}_{\alpha}= \sum_{k\alpha, \sigma} \varepsilon_{k{\alpha}}
c^{\dagger}_{k{\alpha},\sigma} c_{k{\alpha},\sigma} $, corresponds to the
unpolarized reservoirs. The molecule is
described by the simple model
\begin{equation}\label{01m}
  {\cal H}_{M}=\sum_{\sigma=\pm} \varepsilon_{\sigma} d^{\dagger}_{\sigma} d_{\sigma},
\end{equation}
with the energy of two spin orientations  separated by the Zeeman splitting $\varepsilon_{\pm}= \pm  \omega_z$/2.
 The last term of ${\cal
  H}$ is the coupling to the reservoirs,
 \beq{02}
{\cal H}_{M,\alpha}=w  \sum_{k{\alpha},\sigma,\sigma'}
\left[c^{\dagger}_{k{\alpha},\sigma}
  U_{\sigma,\sigma^{\prime}}^{\alpha\dagger}d_{\sigma^{\prime}}+ H.c \right],
\eeq
where $U^\alpha$ are unitary matrices for $\alpha=T,S$.

Let us now notice that the above Hamiltonian does not contain the minimum
ingredients to describe the ESR-STM effect. In fact, the unitary
transformation
$c_{k\alpha,\sigma}^{\prime}=\sum_{\sigma^{\prime}}U_{\sigma,\sigma^{\prime}}^{\alpha}c_{k\alpha,\sigma^{\prime}}$
diagonalizes in spin the tunneling Hamiltonian while leaves unchanged the
Hamiltonians of the unpolarized reservoirs. The result is ${\cal
  H}_{M,\alpha}= w   \sum_{k{\alpha},\sigma} [c^{\prime
  \dagger}_{k{\alpha},\sigma} d_{\sigma}+ H.c]$. Hence, the full Hamiltonian
is a sum over two decoupled spin states and observables such as current noise
cannot present any feature depending on the level spacing in the molecule, as
previously noted in a detailed calculation \cite{gurvitz}.

The next ingredient to explore is a molecular electric dipole $e{\bf r}$ that couples  to an electric field ${\bf E}$, e.g. due to tip proximity \cite{levitov}. Assuming a molecule with several levels,   in the absence of ${\bf E}$, ${\cal H}_M$ has eigenstates $|m,\sigma\rangle$ with eigenvalues
$E_m \pm\half\omega_{z,m}$ ($m=0$ is the ground state) where $\omega_{z,m}$ is the splitting due to a Zeeman term ${\cal H}_Z$. Extending the one-dimensional model \cite{levitov}, the
oscillating electric dipole $\sim \eexp{-i\omega_{z,0}t}$ can be described by the 2nd order matrix element
\beq{03}
\sum_{m\neq 0}\frac{\langle 0, -|{\bf E}\cdot{\bf r}|m,-\rangle\langle
  m,-|{\cal H}_Z|0,+\rangle}{E_0-E_m},
\eeq
where the asymmetry of the molecule allows for $\langle m,-|{\cal
  H}_Z|0,+\rangle\neq 0$ for $m\neq 0$. Adding this off-diagonal term to
 ${\cal H}_M$ results in eigenstates related to $|0,\pm\rangle$, i.e. $d_\sigma$, by a unitary spin rotation.
Therefore Eqs. (\ref{01m}) and (\ref{e02}) retain their structure,  the spin
states decouple and no ESR-STM effect, hence some additional modulation of the tunnel barrier is needed \cite{levitov}.

Then, we keep considering the simple Hamiltonian (\ref{01m}) for the molecule
and we proceed to introduce the main feature of our model. Namely, a direct
tunneling between the tip and substrate so that the Hamiltonian (\ref{e01})
acquires an additional term
\beq{04}
{\cal H}_{TS}=W\sum_{kT,k^{\prime}S,\sigma,\sigma^{\prime}}\left[c_{kT,\sigma}^\dagger U_{\sigma,\sigma^{\prime}}^{TS}c_{k^{\prime}S,\sigma^{\prime}}+H.c.\right]
\eeq
In terms of the operators $c^{\prime}_{k\alpha,\sigma}$ previously defined,
${\cal H}_{TS}$ involves the unitary matrix ${\tilde
  U}=U^TU^{TS}U^{S\dagger}$ which can be written as a general rotation of the
form ${\tilde U}=U^{\rm SO}{U}^S$ with
%\beq{05}
$U^{\rm SO}=\eexp{i\sigma_z\phi_1/2}$,
%\qquad
and $U^{S}=\eexp{i\sigma_z\phi_2/2}\eexp{i\sigma_y\theta/2}\eexp{-i\sigma_z\phi_2/2}$,
%\eeq
where $\sigma_{x,y,z}$ the Pauli matrices. We note that with $\theta=0$ the
spin states decouple, hence $\theta\neq 0$ is essential for the ESR effect.
In terms of the unitary transformation $c^{\prime
  \prime}_{kT,\sigma}=\sum_{\sigma^{\prime}}{\hat
  U}^S_{\sigma,\sigma^{\prime}}c^{\prime}_{kS,\sigma^{\prime}}$ the direct
$TS$ coupling becomes spin diagonal with the matrix $U^{SO}$ while ${\cal
  H}_{M,S}$ becomes off diagonal with ${\hat U}^S$.  The phase $\phi_2$ can be
eliminated by defining $d_\sigma=\eexp{i\sigma\phi_2/2}{\tilde d}_\sigma$,
$c^{\prime}_{kT,\sigma}=\eexp{i\sigma\phi_2/2}{\tilde c}_{kS,\sigma}$ and
$c^{\prime \prime}_{kS,\sigma}=\eexp{i\sigma\phi_2/2}{\tilde
  c}_{kS,\sigma}$. The different Hamiltonian terms become
\beq{06}
 {\cal H}_{TS}&=&W\sum_{kT,k^{\prime} S,\sigma}[{\tilde c}^{\dagger}_{kT,\sigma}\eexp{i\sigma\phi_1}{\tilde c}_{k^{\prime}S,\sigma}
+H.c.]\nonumber\\
{\cal H}_{M,S}&=&w\sum_{kS,\sigma}[{\tilde c}^\dagger_{kS,\sigma}{\tilde d}_\sigma+H.c.]\nonumber\\
{\cal H}_{M,T}&=&w\sum_{kT,\sigma,\sigma^{\prime}}[{\tilde c}^\dagger_{kT,\sigma}\left(\eexp{i\sigma_y\theta/2}\right)_{\sigma,\sigma^{\prime}}{\tilde d}_{\sigma^{\prime}}+H.c.]
\eeq
while the ${\cal H}_T,{\cal H}_S$ terms of (\ref{e01}) remain unchanged with the ${\tilde
  c}_{k\alpha,\sigma},{\tilde d}_\sigma$ operators.

 For this model, the charge current operator flowing into the tip reads
 $\hat{J}_T(t)= \hat{J}_{M \rightarrow T}(t) + \hat{J}_{S \rightarrow T}(t)$,
 where the first term corresponds to the current flowing through the molecule,
 while the second one is due to the direct tunneling $W$. We expect the T-M
 capacitance to be smaller than the M-S one \cite{birk},
 hence the circuit current \cite{but} is dominated by the tip current $\hat{J}_T$.
The corresponding noise spectrum can be decomposed as ${\cal
   S}(t,\omega)= {\cal S}_M(t,\omega)+ {\cal S}_W(t,\omega) $. The first term
 corresponds to current-current correlation functions involving current
 operators through the molecule, while the second one corresponds to those due to
   the direct current between tip and substrate,
 \beq{08}
 {\cal S}_M(t,\omega) &= &  \int_{-\infty}^{+\infty} d \tau \langle  \hat{J}_T(t)  \hat{J}_{M \rightarrow T}(t-\tau)  +\nonumber \\
 & &  \hat{J}_{M \rightarrow T}(t)  \hat{J}_T(t-\tau) \rangle e^{i \omega \tau}, \nonumber \\
 {\cal S}_W(t, \omega) &= & 2 \int_{-\infty}^{+\infty} d \tau \langle
 \hat{J}_{S \rightarrow T}(t) \hat{J}_{S \rightarrow T}(t-\tau) \rangle e^{i
   \omega \tau}.
\eeq
In what follows we focus on the first term, which contains the relevant
information involving the scattering processes through the molecule. The
second one represents a background signal, which is usually subtracted from
the experimental data.  The current is induced by applying a dc
  bias voltage $V$, which relates the chemical potentials of the
  tip and substrate as $\mu_T=\mu_S+eV$. In this stationary case ${\cal S}_{l}(t,\omega) \equiv {\cal S}_l(\omega)$, $l=W,M$.

Consider now the experimentally relevant parameters. The DC current $I= \langle \hat{J}_T \rangle \sim 0.1-1$nA for $V \approx 1$V \cite{review}.  In our model we find
that the Larmor frequency appears in the noise when $W\gg w$, hence the DC conductance is dominated by the $W$ term. The latter is
\cite{ferrer} $G_W=\frac{2e^2}{h}\frac{4x}{(1+x)^2}$,  where $x=\pi^2W^2N_T N_S $ and $N_{T,S}$ are the tip and substrate density of states, taken as constants in the $(\mu_T,\mu_S)$ range. Assuming $N_{T,S} \sim 1/b$, where $b$ is the bandwidth of these reservoirs,
we estimate $W/b\approx 10^{-3}$ corresponding to a weak tunneling regime with $x\ll 1$.
The resonance is sharp, with a width of \cite{review} $\Delta\omega/2\pi\approx 1$MHz, much smaller than the resonance frequency, typically $\omega_z/2\pi\approx 500$MHz. A golden rule estimate, consistent with our data for small $w,W$, gives a resonance width
$\Delta\omega=2\pi w^2/b$. The ensuing estimate for the DC current via the molecule is $\langle J_{M\rightarrow T}\rangle=2e\Delta\omega/2\pi$.
The $\Delta\omega$ value above yields $\sim 10^{-12}$A, verifying that the DC current is indeed dominated by the direct conductance $G_W$.
Considering $b\approx 5 eV$ we estimate  $w/b\approx 10^{-5}$.
We note that the experiments were at room temperature,
$T\gg \omega_z$, hence the linewidth is not sensitive to
temperature. We assume that the chemical potential of the substrate $\mu_S=0$ lies
between the two levels. Our results depend weakly on the position of these levels, as long as at least one of them is in between $\mu_S$ and $\mu_R$.

\begin{figure}[t]
%[htb]
%\begin{center}
\includegraphics[height=0.45\columnwidth,clip]{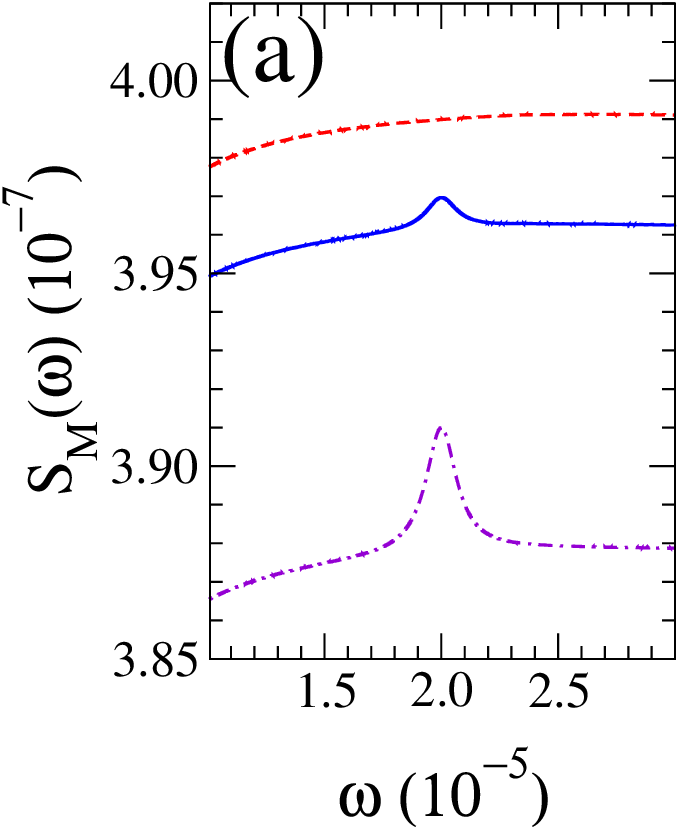}
\hspace{0.05\columnwidth}
\includegraphics[height=0.45\columnwidth,clip]{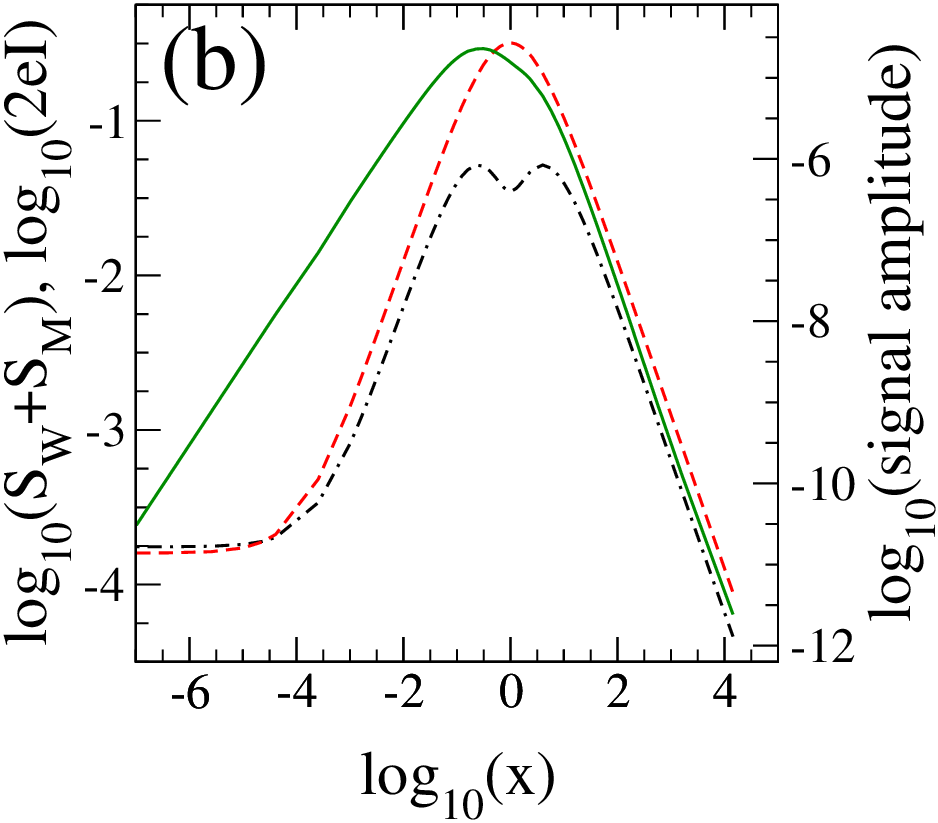}
%\end{center}
\caption{(Color online) (a) noise spectrum ${\cal S}_M$ for various direct
  couplings $W$ between reservoirs: 0 (dashed red), 0.05 (solid blue) and 0.1
  (dash-dotted violet). Parameters are: $\mu_T=0.5$, $\mu_T=0$, $T=0.05$, $w=5
  \cdot 10^{-4}$, $\epsilon_{\pm}=\pm 10^{-5}$ and b=5.  (b) amplitude of the
  signal at the Larmor frequency (solid green), the background noise ${\cal
    S}_W+{\cal S}_M$ (dot-dashed black) and $2eI$ (dashed red) as functions of
  $x$. Parameters are $w=0.01$, $\epsilon_{\pm}=\pm 5 \cdot 10^{-3}$, other
  parameters as in (a).  }\label{fig-W}
\end{figure}

\begin{figure}[t]
%[htb]
%\begin{center}
\includegraphics[height=0.45\columnwidth,clip]{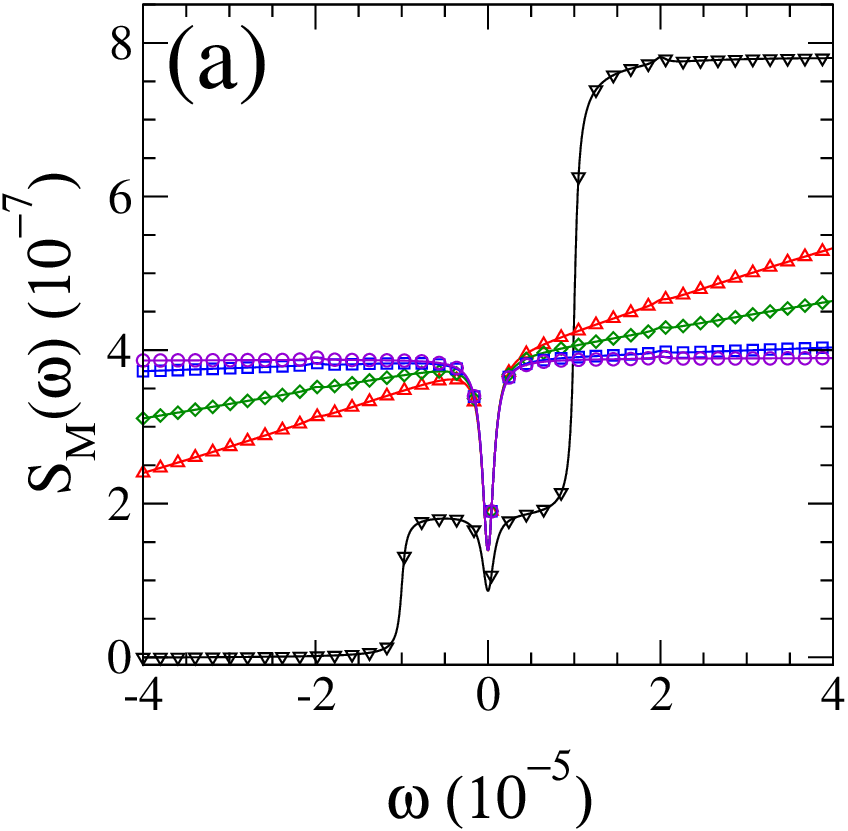}
\includegraphics[height=0.45\columnwidth,clip]{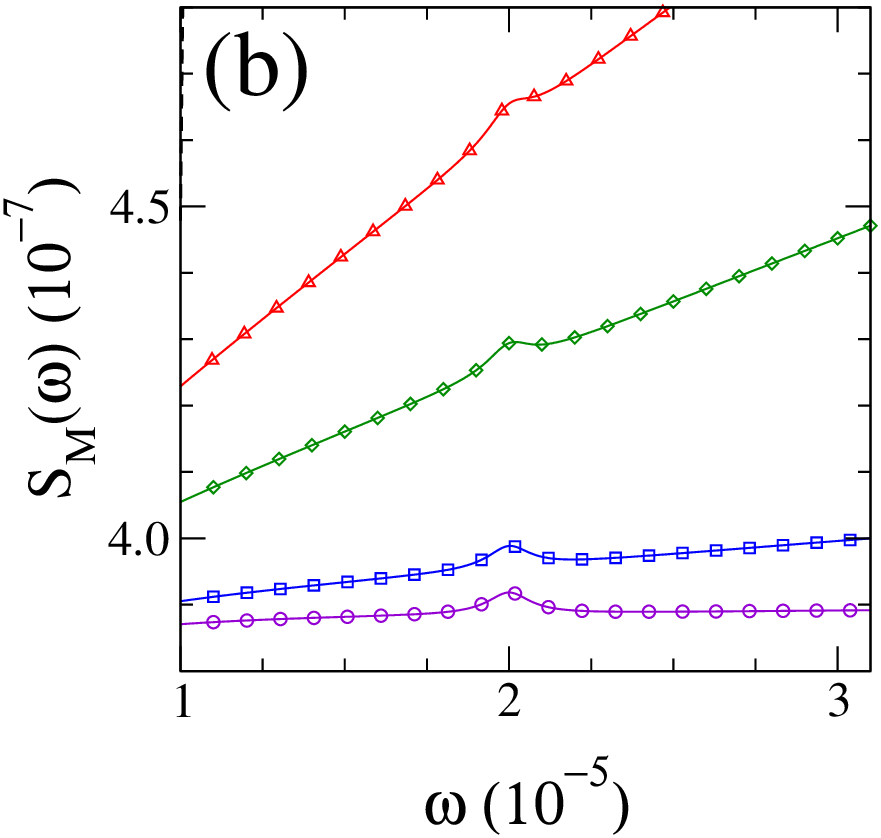}
%\end{center}
\caption{(Color online) Noise spectrum ${\cal S}_M$ at various temperatures
  $T$: 0 (black triangles down), $5 \cdot 10^{-5}$ (red triangles up),
  $10^{-4} $ (green diamond), $5 \cdot 10^{-4}$ (blue squares), and $5 \cdot
  10^{-3}$ (violet circles). Parameters are: $W=0.1 $ and those in
  Fig \ref{fig-W}a.
   (b) shows a magnified part of (a) in the resonance vicinity.}
\end{figure}

We now show results for the noise spectrum in our model. We have calculated ${\cal S}_M(\omega)$ and ${\cal S}_T(\omega)$ following
the procedure of Ref. \onlinecite{noise0}.
We use $\theta=\pi$ and $\phi_1=\pi/2$ to maximize the resonance effect, (the
value of $\phi_1$ is irrelevant to the resonance effect).  We chose $W,w,T,eV,\epsilon_\pm$ not too far from the experimental estimates (in eV units). 
 Both reservoirs are modeled as semi-infinite 1D chains with bandwidth $b$ so that $x=(4 W/b)^2$.

 Fig. \ref{fig-W}a shows the
onset of the Larmor frequency in the noise ${\cal S}_M$ as the direct coupling
$W$ is turned on. We have studied the effect of $w$ on this resonance and extended the model to different tunneling coupling  to the tip and substrate $w_S$ and $w_T$, respectively. We found that the
  the width (as expected above for $w_T=w_S$) as well as the amplitude scale as $w_S w_T$.
In Fig. \ref{fig-W}b we show the amplitude of the signal as
a function of $x$, the background noise and  the classical shot noise $2eI$. The background noise is taken as the mean in the range $\omega_z\pm0.1\omega_z$, where the noise varies by less than $1\%$. We note that
for $x\ll 1$ or $x\gg 1$ the background is the classical shot noise ${\cal S}_W+{\cal S}_M\approx 2eI$.
 The current ratio is $\langle J_{M\rightarrow T}\rangle/\langle J_{S\rightarrow T}\rangle\approx 4\pi w^2/(bxeV)$ (for $x\ll 1$) which for the parameters of Fig. \ref{fig-W}b is $\approx 10^{-4}/x$. Hence at $x<10^{-4}$ the current, as well as the background noise, is dominated by that via the molecule and becomes $x$ independent.
  The case $x=1$ corresponds to perfect ballistic matching between the reservoirs, accounting for the maxima in the figure. 

Fig. \ref{fig-W}b shows that the signal intensity at $x\ll 1$ is linear with $x$, same as ${\cal S}_W\sim G_W\sim x$. The signal to background ratio at $x=0.01$   is $\sim 10^{-3}$. Scaling to $w=5\cdot 10^{-5}$ reduces this ratio to $\sim 10^{-8}$. For this $w$ the background is $S_W$ dominates, hence the ratio above is $x$ independent and applies to the experimental situation.

The temperature dependence is shown in Fig. 2. The resonance itself is not
sensitive to temperature (see Fig. 2b), consistent with the
experimental data \cite{review}. In the case of $T=0$ there are features of
the noise spectrum (steps) at the frequencies corresponding to the molecular
levels $\varepsilon_{\pm}$, which are typical of the noise spectrum of
two-level systems \cite{rothstein}. These features are washed up by thermal
fluctuations as soon as $T> \omega_z$. The dip at $\omega=0$ is similar to the one seen in Ref. \onlinecite{rothstein},  with the latter due to charge conservation at the molecule.

\begin{figure}[htb]
%[htb]
%\begin{center}
\includegraphics[height=0.45\columnwidth,clip]{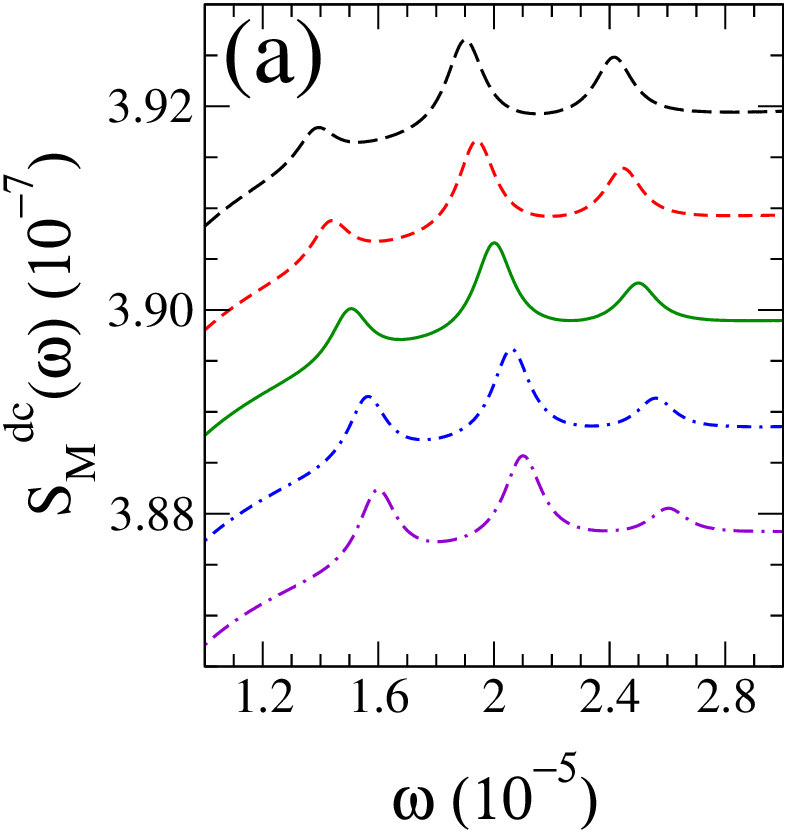}
\hspace{0.01\columnwidth}
\includegraphics[height=0.45\columnwidth,clip]{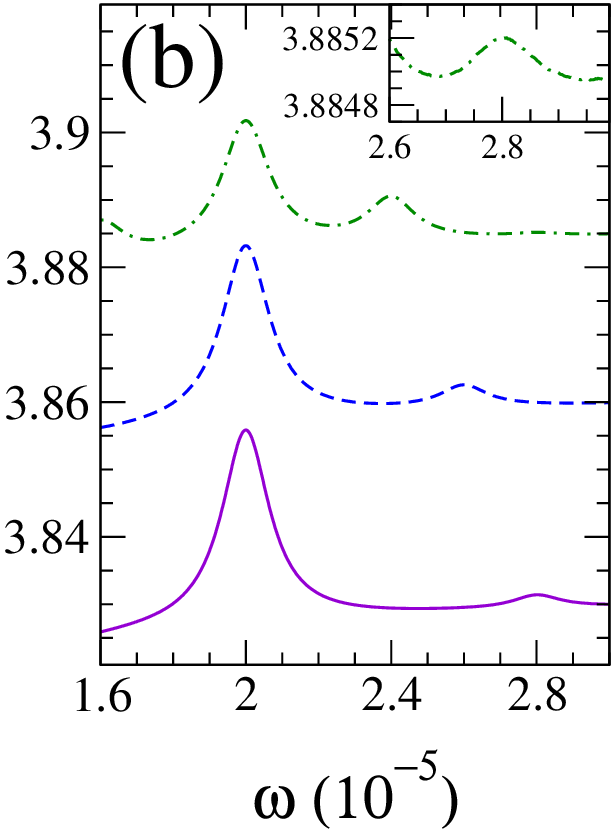}
%\end{center}
\caption{(Color online) (a) noise spectrum ${\cal S}^{\rm dc}_M$ for
  perpendicular ac field amplitude $\Delta_{\perp}=5\cdot 10^{-6}$ and various
  frequencies $\Omega_0$: $1.9\cdot 10^{-5}$ (long-dashed black), $1.94 \cdot
  10^{-5}$ (dashed red), $2 \cdot 10^{-5}$ [resonant] (solid green), $2.06
  \cdot 10^{-5}$ (dashed-dotted blue), and $2.1 \cdot 10^{-5}$
  (long-dashed-dotted violet). (b) noise spectrum ${\cal S}^{\rm dc}_M$ for
  parallel ac field amplitude $\Delta_{\parallel} = 2 \cdot 10^{-6}$ and
  various frequencies $\Omega_0$: $4 \cdot 10^{-6}$ (dashed-dotted green), $6
  \cdot 10^{-6}$ (dashed blue), and $8 \cdot 10^{-6}$ (solid violet). Inset:
  second harmonic corresponding to $\Omega_0 = 4 \cdot 10^{-6}$. Parameters
  are: $W=0.1$ and those in Fig. 1a. The curves below the top one in (a) and
  (b) are displaced in height for clarity.}
\end{figure}

In Fig. 3a we show the noise in presence of an additional ac
magnetic field perpendicularly applied with respect to ${\bf B}$,
$B_{\perp}(t)=\Delta_{\perp} \cos(\Omega_0 t)\sigma_1$ that induces
transitions between the molecule levels. For the isolated molecule the
transition amplitude at time $t$ is
\beq{09}
\langle -,t|+,t=0\rangle=\frac{\Delta_{\perp}}{a}\eexp{-\half i\Omega_0  t}\sin(\half a t)
 \eeq
 where $a=\sqrt{(\Omega_0-\omega_z)^2+\Delta_{\perp}^2}$ is the Rabi frequency. In the presence of
 $B_{\perp}(t)$, the current-current correlation functions (\ref{e08}) depend
 on the time argument $t$. In what follows we consider the ensuing dc
 components
 \beq{10}
  {\cal S}^{\rm dc}_l(\omega) = \frac{\Omega_0}{2 \pi} \int_0^{\Omega_0/(2\pi)} dt {\cal S}_l(t,\omega).
 \eeq
 In the present case, we resort to the non-equilibrium Green
 function formalism of Refs. \onlinecite{noiset} to evaluate this function.
 Fig. 3a shows splitting of the line into three signals located at $\Omega_0$ and
 $\Omega_0 \pm a$. The amplitudes of the signals increase as the satellite
 peaks become closer to $\omega_z$. In usual ESR one usually observes the average spin with a resonance at $\omega_z$ while the amplitude of the Rabi frequency $a$ decays by a $T_2$ process.
 The analog of our current correlation is, however, the spin-spin correlation
 in ESR. The latter is in fact related to light scattering from a 2-level
 system showing a ``Mollow triplet'' \cite{muller,ulrich}.

 In Fig. 3b we show the noise in presence of an ac magnetic
 field parallel to ${\bf B}$, $B_{\parallel}(t) = \Delta_{\parallel}
 \cos(\Omega_0 t)\sigma_z$ which produces an oscillation in the energy levels
 around their equilibrium values with amplitude $ \Delta_{\parallel} $ and
 frequency $\Omega_0$. In this case, the time evolution of the off diagonal
 $\sigma_1$ element is
\beq{11}
e^{i \int_0^t (\omega_z + \Delta_{\parallel} \cos(\Omega_0 t')) dt'} = e^{i \omega_z t}
  \sum_{n=-\infty}^{\infty} J_n \left(\frac{\Delta_{\parallel}}{\Omega_0}\right) e^{i n
    \Omega_0 t}
\eeq
where $J_n(x)$ are the Bessel functions. For $\Delta_\parallel \ll \Omega_0$
only the first terms of the series with $n=0,\pm 1$ contribute being the term
with $n=0$ with $J_0(0)=1$ the dominant one. Thus, in remarkable contrast with
the case of a perpendicular ac field, the strongest noise signal appears at
$\omega_z$ with two sizable satellites at $\omega_z \pm \Omega_0$. This is
illustrated in Fig. 3b, where the main peak and the right satellite are shown
for several values of $\Omega_0$; the inset zooms on the weaker second order
order satellite at $\omega_z + 2 \Omega_0$. A case with many sidebands was in
fact studied by ERS-STM \cite{manassen3} and is consistent with
Eq. (\ref{e11}).  It is remarkable that these well known features from ESR are
reproduced in the current noise spectra.

{\em Discussion:} Our model assumes that the spin-orbit interaction is
significant, for at least one of the tunneling terms. We note that the strong
electric field near the tip can enhance the spin-orbit coupling in the
molecule.  We consider now several unusual features of the data that our model
can account for: (i) A sharp resonance even at high temperatures $T\gg
\omega_z$, (ii) insensitivity to the details of the spin defect, i.e. to the
positions of its levels between the tip and substrate chemical potentials,
(iii) contour plots \cite{manassen1,manassen2} showing that the signal is
maximal at $\sim 1$nm from a center, hence a significant direct coupling $W$
bypassing the spin can be achieved. (iv) We account for the ESR-STM phenomenon
with unpolarized tip or substrate, in contrast with previous models
\cite{bulaevskii,gurvitz} that require polarized leads.

We note that the background noise is not measured in the experiment since the
modulation technique \cite{review} measures the derivative of the noise
spectra. Furthermore, the signal intensity is under study \cite{manassen4} as
it is highly sensitive to uncertainties in the feedback and impedance matching
circuits.  We estimate that the signal to background intensity is $\sim
10^{-8}$ as discussed above.  Furthermore, we predict the appearance of
triplet lines when adding a time dependent field perpendicular to the DC
one. The spacing of these lines is determined by the Rabi frequency. We also
predict multiple sidebands for modulation with parallel field, partly seen in
experiment \cite{manassen3}. In conclusion, our model presents a solution to a
long standing puzzle, paving the way for more controlled single spin detection
via ESR-STM.

\begin{acknowledgments}
  We thank for stimulating discussions with Y. Manassen, A. Golub,
  S. A. Gurvitz, A. Janossy, L. S. Levitov, I. Martin, M. Y. Simmons,
  F. Simon, E. I. Rashba, S. Rogge, E. A. Rothstein, A. Shnirman,
  G. Z\'{a}rand and A. Yazdani. This research was supported by THE ISRAEL
  SCIENCE FOUNDATION (BIKURA) (grant No. 1302/11), by the Israel-Taiwanese
  Scientific Research Cooperation of the Israeli Ministry of Science and
  Technology (BH), as well as CONICET, MINCyT and UBACyT from Argentina (LA
  and AC).
\end{acknowledgments}

\end{document}